# Reply to 'Comment on "Relativistic extension of shape-invariant potentials"'

A. D. Alhaidari

Physics Department, King Fahd University of Petroleum & Minerals, Box 5047, Dhahran 31261, Saudi Arabia

E-mail: **haidari@mailaps.org**

We concur with de Castro's observation that the gauge considerations of our approach are not valid. Nevertheless, except for an error that will be corrected, all of our findings are accurate independent of those considerations.

The Hamiltonian that resulted in the radial equation (1) of our Paper [1] is not the minimum coupling Hamiltonian $H$ shown on the second page of the Paper but the one obtained from it by replacing the two off-diagonal terms $\alpha\vec{\sigma}\cdot\vec{A}$ with $\pm i\alpha\vec{\sigma}\cdot\vec{A}$, respectively. Consequently, our interpretation of $(V,\hat{r}W)$ as the electromagnetic potential and the statement that "$W(r)$ is a gauge field" are not correct. Likewise, calling equation (3) in the Paper, or any other derived from it, as the "gauge fixing condition" is not accurate. This has to be replaced everywhere by the term "constraint". Therefore, it is justifiable to conclude, as de Castro did [2], that the gauge considerations presented in our Paper are invalid. Nonetheless, except for an error which is corrected below, all developments based on, and findings subsequent to equation (1) still stand independent of those considerations.

The error, which was pointed out in the Comment of assigning the inadmissible value $\kappa = 0$ in the cases where $V(r) = 0$, was hastily made to eliminate the centrifugal barrier $\kappa(\kappa+1)/r^2$ and the term $2\kappa W/r$ simultaneously from equation (11) so that we end up with the "super-potential" $W^2 - W'$. This mistake, which will now be corrected, affects only the Dirac-Rosen-Mörse II, Dirac-Scarf, and Dirac-Pöschl-Teller problems. Eliminating these two terms can be achieved properly by replacing the potential function $W(r)$ given in the Paper for each of the three problems by $W(r) - \kappa/r$, where $\kappa$ is now arbitrary. That is, in equations (10) and (11) of the Paper and in the Table we substitute the following potential function for the corresponding problem:

Dirac-Rosen-Mörse II: $W(r) = F\coth(\lambda r) - G\operatorname{csch}(\lambda r) - \kappa/r$

Dirac-Scarf: $W(r) = F\tanh(\lambda r) + G\operatorname{sech}(\lambda r) - \kappa/r$

Dirac-Pöschl-Teller: $W(r) = F\tanh(\lambda r) - G\coth(\lambda r) - \kappa/r$

where $F$, $G$, and $\lambda$ are the potential parameters defined in the paper. This gives the same differential equations for the spinor components and reproduces the same solutions (energy spectrum and wave functions) as those given in the Paper for each of the three problems.